*Gene Expression*

# CplexA: a *Mathematica* package to study macromolecular-assembly control of gene expression

Jose M. G. Vilar[1,2,*] and Leonor Saiz[3,*]

[1]Biophysics Unit (CSIC-UPV/EHU) and Department of Biochemistry and Molecular Biology, University of the Basque Country, P.O. Box 644, 48080 Bilbao, Spain

[2]IKERBASQUE, Basque Foundation for Science, 48011 Bilbao, Spain

[3]Department of Biomedical Engineering, University of California, 451 E. Health Sciences Drive, Davis, CA 95616, USA

**ABSTRACT**
**Summary:** Macromolecular assembly vertebrates essential cellular processes, such as gene regulation and signal transduction. A major challenge for conventional computational methods to study these processes is tackling the exponential increase of the number of configurational states with the number of components. CplexA is a *Mathematica* package that uses functional programming to efficiently compute probabilities and average properties over such exponentially large number of states from the energetics of the interactions. The package is particularly suited to study gene expression at complex promoters controlled by multiple, local and distal, DNA binding sites for transcription factors.
**Availability:** CplexA is freely available together with documentation at http://sourceforge.net/projects/cplexa/.
**Contact:** j.vilar@ikerbasque.org; lsaiz@ucdavis.edu
**Supplementary Information:** Supplementary data are available at *Bioinformatics* online.

## 1 INTRODUCTION

The study of the cellular behavior from the molecular components often requires approximations in terms of chemical reactions. There are, however, many instances, such as combinatorial macromolecular assembly, that cannot be efficiently described in terms of chemical reactions (Saiz and Vilar, 2006). Macromolecular complexes are typically made of smaller building blocks with a modular organization that can be combined in a number of different ways. The result of each combination is a specific molecular species. Therefore, there are potentially as many reactants as the number of possible ways of arranging the different elements, which grows exponentially with the number of the constituent elements.

Several approaches have been developed to tackle this exponentially large multiplicity in the number of states. They involve a diversity of methodologies that range from stochastic configuration sampling (Le Novere and Shimizu, 2001; Saiz and Vilar, 2006) to automatic generation of all the underlying equations (Hlavacek, et al., 2006). The complexity of the general problem makes each of these approaches work efficiently only on a particular type of problems, be it conformational changes, multi-site phosphorylation, or oligomerization (Borisov, et al., 2006; Bray and Lay, 1997; Saiz and Vilar, 2006).

The package CplexA focuses on macromolecular assembly on a template. The prototypical example is a complex promoter where DNA provides a flexible template for the assembly of transcription factors. CplexA provides mathematical tools to infer the probability of having a given set of configurations. In the case of a promoter, it would be the probability of having a pattern of transcription factors bound, which can be used to infer the resulting transcription rate in a way that can be integrated with other software to study the dynamics of cellular networks (Shapiro, et al., 2003).

This type of systems has traditionally been studied by writing a table with entries for each state and the corresponding free energies and associated probabilities, which are used to compute average quantities such as effective transcription rates (Ackers, et al., 1982). As the number of states increases exponentially, the approach becomes impracticable. In this type of systems, however, it is possible to take advantage of the unambiguous structures that macromolecular complexes typically have on a template and use "table-centric" equivalent mathematical approaches that are able to capture this complexity in simple terms (Saiz and Vilar, 2006).

## 2 METHODS

The mathematical approach underlying CplexA is discussed in detail in (Saiz and Vilar, 2006). It specifies the system by a set of $N$ state variables, $S = \{s_1, ... s_i, ... s_N\}$, that can be either 1 to indicate that a property is present (e.g. binding or conformation) or 0 to indicate that it is not. The free energy, $\Delta G(S)$, and a configuration pattern, $\Gamma(S)$, can generally be expressed as a function of these state variables. The probability of the configuration pattern is obtained from

---

*To whom correspondence should be addressed.

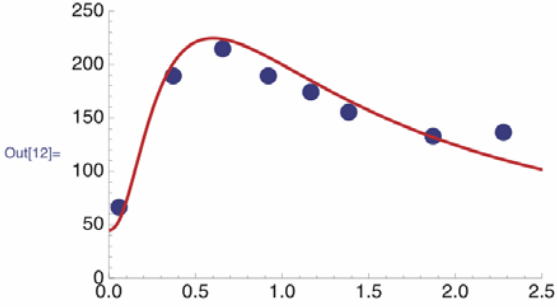

```
In[8]:=  S:={sOL1, sOL2, sOL3, sOR1, sOR2, sOR3, sL}

In[9]:=  ΔGλ:=sOR1 (-12.7-RT Log[n])+sOR2 (-10.7-RT Log[n])+
         sOR3 (-10.2-RT Log[n])-3 sOR1 sOR2-3 sOR2 sOR3+
         3 sOR1 sOR2 sOR3 + sOL1 (-13.8-RT Log[n])+
         sOL2 (-12.1-RT Log[n])+sOL3 (-12.4-RT Log[n])-
         2.5 sOL1 sOL2-2.5 sOL2 sOL3+2.5 sOL1 sOL2 sOL3+
         sL (21-21.2 sOL1 sOL2 sOR1 sOR2-3 sOL3 sOR3);

In[10]:= Γrm:=
         (1-sOR3) (45+((240-45) sL+(460-45) (1-sL)) sOR2)

In[11]:= AΓrm=AvConf[Γrm,ΔGλ,S];

In[12]:= Show[plotPRM,Plot[AΓrm/.{RT->0.6,n->NF[nn]},
         {nn,0,2.5},PlotStyle->{Red,Thick}]]
```

Out[12]:=

**Figure 1.** Use of CplexA with *Mathematica 7* to compute the effective transcription rate at the $P_{RM}$ promoter of Phage λ. The graph in line "Out[12]" shows the computed average transcription (solid red line) as a function of the normalized CI monomer concentration. The filled circles correspond the experimentally measured activity of the $P_{RM}$ promoter (Dodd, et al., 2004). (For details see Supplementary Information.)

$$\overline{\Gamma} = \frac{\sum_S \Gamma(S) e^{-\Delta G(S)/RT}}{\sum_S e^{-\Delta G(S)/RT}} \quad (1)$$

by computing the thermodynamic average over all $2^N$ possible values of $S$.

## 3 APPLICATION

The package CplexA provides the function AvConf[ $\Gamma$ , $\Delta G$ , $S$ ] that computes the thermodynamic average $\overline{\Gamma}$. Figure 1 illustrates the use of CplexA with *Mathematica 7* to compute the effective transcription rate at the $P_{RM}$ promoter of Phage λ (Saiz and Vilar, 2006). This system consists of two sets of three contiguous binding sites for the CI dimer. The two sets, known as left and right operators, are 2kb apart from each other. CI dimers bound at different operators can interact with each other by looping the intervening DNA. In this case, just a few lines of code, from lines "In[8]" to "In[11]" in Figure 1, can achieve the same results as a table with entries for each of the 128 states. The state of the system, $S$, is described by six binding and one looping state variables in line "In[8]". The free energy, $\Delta G(S)$, in line "In[9]" includes, in a very compact manner, binding to each of the six sites as a function of the dimer concentration, interactions between neighboring dimers, DNA looping, and the formation of octamers and tetramers between dimers bound at different sets of binding sites. The transcription rate, $\Gamma(S)$, as a function of the binding and looping state is given in line "In[10]". Its average value, $\overline{\Gamma}$=AvConf[ $\Gamma$ , $\Delta G$ , $S$ ] , closely matches the experimental data on the transcriptional activity of the promoter (Dodd, et al., 2004).

CplexA also provides the function DGTable[ $\Delta G$ , $S$ ], which constructs a table with the free energy and statistical weight (Boltzmann factor) of each state that has a non-zero probability.

## 4 IMPLEMENTATION

The critical issue in the implementation of the function AvConf[ $\Gamma$ , $\Delta G$ , $S$ ] is dealing with the combinatorial explosion in the number of states. Using state variables overcomes the combinatorial explosion in the specification of the problem but not in the sum over all the states, which still grows as $2^N$. A fundamental advantage of using a computer algebra system, such as *Mathematica*, over imperative programming languages, such as Fortran, C, or Java, is that it allows for the direct manipulation of functions. In CplexA, the implementation of the sum over all possible values of $S$ in the numerator and denominator of equation (1) is performed in $N$ steps, rather than in $2^N$ , by using the backwards recursion

$$f_{N-1}(s_1,...s_{N-1}) = f_N(s_1,...s_{N-1},0) + f_N(s_1,...s_{N-1},1) \ .$$

Starting this recursion with the functions

$$f_N(s_1,...s_N) \equiv \Gamma(S) e^{-\Delta G(S)/RT} \ \text{ and } \ f_N(s_1,...s_N) \equiv e^{-\Delta G(S)/RT}$$

leads to the sought values of the sums as

$$\sum_S \Gamma(S) e^{-\Delta G(S)/RT} = f_0 \ \text{ and } \ \sum_S e^{-\Delta G(S)/RT} = f_0 \ ,$$

respectively, after the $N$ steps of the recursion have been performed. With this method, the actual computational complexity depends on the specific form of $f_N(s_1,...s_N)$ and does not necessarily increase proportionally to the number of states. For instance, for a linear array of binding sites with next-neighbors interactions, the CPU time needed to compute the average occupancy for the case of 40 sites is only a factor ~8 higher than that needed for 20 sites, whereas the number of states increases by a factor ~$10^6$ (see Supplementary Information).

## ACKNOWLEDGEMENTS

*Funding*: Ministerio de Ciencia e Innovacion (FIS2009-10352); University of California, Davis.

## REFERENCES

Ackers, G.K., Johnson, A.D. and Shea, M.A. (1982) Quantitative Model for Gene-Regulation by Lambda-Phage Repressor, Proc Natl Acad Sci U S A, 79, 1129-1133.

Borisov, N.M., Markevich, N.I., Hoek, J.B. and Kholodenko, B.N. (2006) Trading the micro-world of combinatorial complexity for the macro-world of protein interaction domains, Biosystems, 83, 152-166.

Bray, D. and Lay, S. (1997) Computer-based analysis of the binding steps in protein complex formation, Proc Natl Acad Sci U S A, 94, 13493-13498.

Dodd, I.B., Shearwin, K.E., Perkins, A.J., Burr, T., Hochschild, A. and Egan, J.B. (2004) Cooperativity in long-range gene regulation by the lambda CI repressor, Genes Dev, 18, 344-354.

Hlavacek, W.S., Faeder, J.R., Blinov, M.L., Posner, R.G., Hucka, M. and Fontana, W. (2006) Rules for modeling signal-transduction systems, Sci STKE, 2006, re6.

Le Novere, N. and Shimizu, T.S. (2001) STOCHSIM: modelling of stochastic biomolecular processes, Bioinformatics, 17, 575-576.

Saiz, L. and Vilar, J.M.G. (2006) Stochastic dynamics of macromolecular-assembly networks, Mol Syst Biol, 2, 2006.0024.

Shapiro, B.E., Levchenko, A., Meyerowitz, E.M., Wold, B.J. and Mjolsness, E.D. (2003) Cellerator: extending a computer algebra system to include biochemical arrows for signal transduction simulations, Bioinformatics, 19, 677-678.

# Supplementary Information for "CplexA: a Mathematica package to study macromolecular-assembly control of gene expression"

*Jose M. G. Vilar* [1,2] *and Leonor Saiz* [3]

[1]*Biophysics Unit (CSIC-UPV/EHU) and Department of Biochemistry and Molecular Biology, University of the Basque Country, P.O. Box 644, 48080 Bilbao, Spain*

[2]*IKERBASQUE, Basque Foundation for Science, 48011 Bilbao, Spain*

[3]*Department of Biomedical Engineering, University of California, 451 East Health Sciences Drive, Davis, CA 95616, USA*

In[1]:= **SetDirectory@NotebookDirectory[];**

In[2]:= **<< CplexA`**

## Detailed calculations for Figure 1

In[3]:= **(\* Experimental data for the transcriptional activity of the PRM and PR promoters of phage lambda from ref. Dodd, I.B.,Shearwin,K.E.,Perkins,A.J.,Burr,T.,Hochschild, A.and Egan, J.B.(2004) Cooperativity in longrange gene regulation by the lambda CI repressor, Genes Dev,18,344-354 \*)**
**PRMdata := {{0.058, 66.367}, {0.373, 189.358}, {0.658, 214.233}, {0.922, 188.897}, {1.168, 173.696}, {1.385, 155.271}, {1.875, 133.160}, {2.281, 136.384}};**
**PRdata := {{0.063, 1057.627}, {0.373, 585.763}, {0.658, 204.746}, {0.921, 70.508}, {1.171, 27.119}, {1.384, 18.983}, {1.875, 12.203}, {2.278, 9.492}};**

In[5]:= **(\* Plot of the experimental data to use later on \*)**
**plotPRM := ListPlot[PRMdata, PlotMarkers → {Automatic, 17}, PlotRange → {{0, 2.5}, {0, 250}}, ImageSize → 300, BaseStyle → {FontFamily → "Helvetica", FontSize → 12}];**
**plotPR := ListPlot[PRdata, PlotMarkers → {Automatic, 17}, PlotRange → {{0, 2.5}, {0, 1200}}, ImageSize → 300, BaseStyle → {FontFamily → "Helvetica", FontSize → 12}];**

In[7]:= **(\* Computes the CI dimer concentration as a function of CI monomer concentrations \*)**
**NF[nn_] := 9.38419 \* 10^-14 + 7.0922 \* 10^-10 nn - 1.15373 \* 10^-11 Sqrt[0.0000661586 + 1. nn]**

In[8]:= **(\* State variables \*)**
**S := {sOL1, sOL2, sOL3, sOR1, sOR2, sOR3, sL}**

In[9]:= **(\* Free energy \*)**
**ΔGλ := sOR1 (-12.7 - RT Log[n]) + sOR2 (-10.7 - RT Log[n]) + sOR3 (-10.2 - RT Log[n]) - 3 sOR1 sOR2 - 3 sOR2 sOR3 + 3 sOR1 sOR2 sOR3 + sOL1 (-13.8 - RT Log[n]) + sOL2 (-12.1 - RT Log[n]) + sOL3 (-12.4 - RT Log[n]) - 2.5 sOL1 sOL2 - 2.5 sOL2 sOL3 + 2.5 sOL1 sOL2 sOL3 + sL (21 - 21.2 sOL1 sOL2 sOR1 sOR2 - 3 sOL3 sOR3);**

In[10]:= **(\* Transcriptional activity of the PRM promoter \*)**
**Γrm := (1 - sOR3) (45 + ((240 - 45) sL + (460 - 45) (1 - sL)) sOR2);**

In[11]:= **(\* Average transcriptional activity of the PRM promoter \*)**
**AΓrm = AvConf[Γrm, ΔGλ, S];**

In[12]:= (* Computed average transcriptional activity of the PRM
 promoter (red line) and its experimental counterpart (blue circles)
 as a function of the normalized CI dimer concentration *)
Show[plotPRM, Plot[AΓrm /. {RT → 0.6, n → NF[nn]}, {nn, 0, 2.5}, PlotStyle → {Red, Thick}]]

Out[12]=

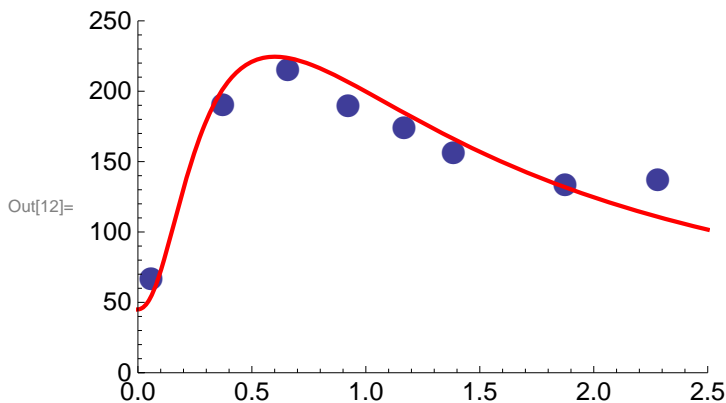

In[13]:= (* Transcriptional activity of the PR promoter *)
Γr := (1 - sOR1) 1200;

In[14]:= (* Average transcriptional activity of the PR promoter *)
AΓr = AvConf[Γr, ΔGλ, S];

In[15]:= (* Computed average transcriptional activity of the PR
 promoter (red line) and its experimental counterpart (blue circles)
 as a function of the normalized CI dimer concentration *)
Show[plotPR, Plot[AΓr /. {RT → 0.6, n → NF[nn]},
 {nn, 0, 2.5}, PlotStyle → {Red, Thick}, PlotRange → All]]

Out[15]=

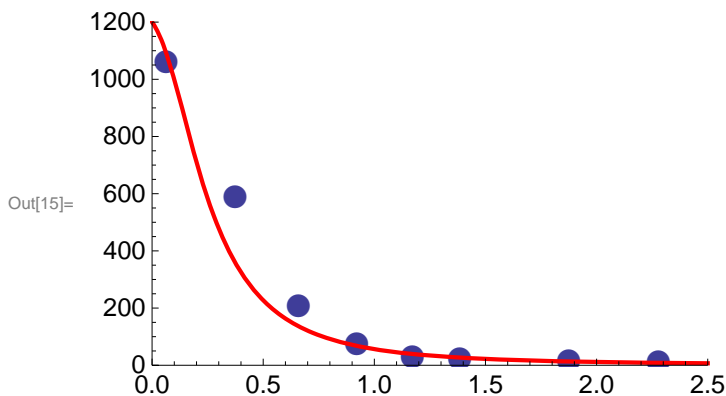

## Sum over an exponentially large number of states in sub-exponential times

In[16]:= (* Definition of a linear array of binding sites with next-
 neighbor interactions for molecules with concentration
 n. The function NNChain returns a list with the state variables,
 the free energy of the chain, the average occupancy,
 and the time needed for the computation. NumSites is the number of sites;
 NNint is the energy of the next-neighbor interaction between bound molecules;
 the free energy of binding is -1; and the thermal energy is 1. *)
NNChain[NumSites_, NNint_] := Module[{S, ΔGB, res},
 S = Table[ToExpression["s" <> ToString[i]], {i, 1, NumSites}];
 ΔGB = Simplify[Sum[(-1 - Log[n]) S[[i]], {i, 1, NumSites}] +
  NNint Sum[S[[i]] S[[i + 1]], {i, 1, NumSites - 1}]];
 res = Timing[AvConf[Mean[S], Simplify[ΔGB], S, 1]]; {S, ΔGB, res[[2]], res[[1]]}]

In[17]:= `(* Analytical expression of the average`
`   occupancy of a linear array of 10 binding sites *)`
`NNChain10 = NNChain[10, NNint][[3]]`

Out[17]= $\Big( e\, n\, \big( 5\, e^{9\,NNint} + 9\, e^{1+8\,NNint}\, n + 36\, e^{1+9\,NNint}\, n + 12\, e^{2+7\,NNint}\, n^2 + 84\, e^{2+8\,NNint}\, n^2 + 84\, e^{2+9\,NNint}\, n^2 + 14\, e^{3+6\,NNint}\, n^3 +$
$126\, e^{3+7\,NNint}\, n^3 + 210\, e^{3+8\,NNint}\, n^3 + 70\, e^{3+9\,NNint}\, n^3 + 15\, e^{4+5\,NNint}\, n^4 + 150\, e^{4+6\,NNint}\, n^4 + 300\, e^{4+7\,NNint}\, n^4 +$
$150\, e^{4+8\,NNint}\, n^4 + 15\, e^{4+9\,NNint}\, n^4 + 15\, e^{5+4\,NNint}\, n^5 + 150\, e^{5+5\,NNint}\, n^5 + 300\, e^{5+6\,NNint}\, n^5 +$
$150\, e^{5+7\,NNint}\, n^5 + 15\, e^{5+8\,NNint}\, n^5 + 14\, e^{6+3\,NNint}\, n^6 + 126\, e^{6+4\,NNint}\, n^6 + 210\, e^{6+5\,NNint}\, n^6 + 70\, e^{6+6\,NNint}\, n^6 +$
$12\, e^{7+2\,NNint}\, n^7 + 84\, e^{7+3\,NNint}\, n^7 + 84\, e^{7+4\,NNint}\, n^7 + 9\, e^{8+NNint}\, n^8 + 36\, e^{8+2\,NNint}\, n^8 + 5\, e^9\, n^9 \big) \Big) \Big/$
$\Big( 5\, \big( e^{9\,NNint} + 10\, e^{1+9\,NNint}\, n + 9\, e^{2+8\,NNint}\, n^2 + 36\, e^{2+9\,NNint}\, n^2 + 8\, e^{3+7\,NNint}\, n^3 + 56\, e^{3+8\,NNint}\, n^3 +$
$56\, e^{3+9\,NNint}\, n^3 + 7\, e^{4+6\,NNint}\, n^4 + 63\, e^{4+7\,NNint}\, n^4 + 105\, e^{4+8\,NNint}\, n^4 + 35\, e^{4+9\,NNint}\, n^4 + 6\, e^{5+5\,NNint}\, n^5 +$
$60\, e^{5+6\,NNint}\, n^5 + 120\, e^{5+7\,NNint}\, n^5 + 60\, e^{5+8\,NNint}\, n^5 + 6\, e^{5+9\,NNint}\, n^5 + 5\, e^{6+4\,NNint}\, n^6 + 50\, e^{6+5\,NNint}\, n^6 +$
$100\, e^{6+6\,NNint}\, n^6 + 50\, e^{6+7\,NNint}\, n^6 + 5\, e^{6+8\,NNint}\, n^6 + 4\, e^{7+3\,NNint}\, n^7 + 36\, e^{7+4\,NNint}\, n^7 + 60\, e^{7+5\,NNint}\, n^7 +$
$20\, e^{7+6\,NNint}\, n^7 + 3\, e^{8+2\,NNint}\, n^8 + 21\, e^{8+3\,NNint}\, n^8 + 21\, e^{8+4\,NNint}\, n^8 + 2\, e^{9+NNint}\, n^9 + 8\, e^{9+2\,NNint}\, n^9 + e^{10}\, n^{10} \big) \Big)$

In[18]:= `(* Plot of the analytical expression of the average`
`   occupancy of a linear array of 10 binding sites as a function of`
`   the energy of interaction and the logarithm of the concentration *)`
`Plot3D[NNChain10 /. n → 10^x, {NNint, -5, 5}, {x, -2, 2},`
`   AxesLabel → {"NNint", "Log[n]", "Occupancy", "", ""}]`

Out[18]=

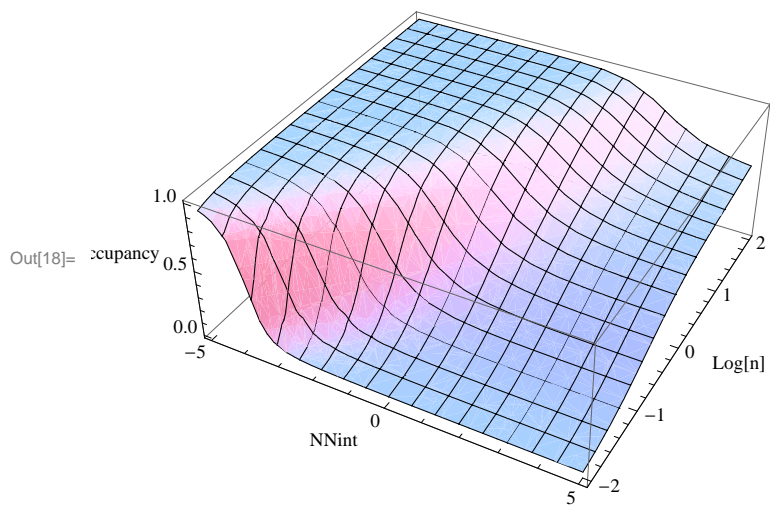

In[19]:= `TimesTab = Table[{NumSites, NNChain[NumSites, -1][[4]]}, {NumSites, 1, 40}]`

Out[19]= {{1, 0.003723}, {2, 0.00872}, {3, 0.016504}, {4, 0.026798}, {5, 0.049121}, {6, 0.067722},
{7, 0.102238}, {8, 0.134266}, {9, 0.179601}, {10, 0.224848}, {11, 0.281501},
{12, 0.234544}, {13, 0.274352}, {14, 0.327549}, {15, 0.403452}, {16, 0.471403},
{17, 0.561451}, {18, 0.653952}, {19, 0.7683}, {20, 0.901277}, {21, 1.02825}, {22, 1.18254},
{23, 1.36579}, {24, 1.49554}, {25, 1.71374}, {26, 1.93023}, {27, 2.19595}, {28, 2.49536},
{29, 2.79143}, {30, 3.1087}, {31, 3.51332}, {32, 3.89292}, {33, 4.33852}, {34, 4.87941},
{35, 5.32834}, {36, 5.93174}, {37, 6.64703}, {38, 7.36409}, {39, 7.94771}, {40, 8.75999}}

In[20]:= `(* The CPU time (in seconds) needed to compute the average occupancy is`
`  plotted as a function of the number of binding sites of the linear array *)`
`ListLogPlot[TimesTab, Joined → True, PlotRange → All]`

Out[20]=

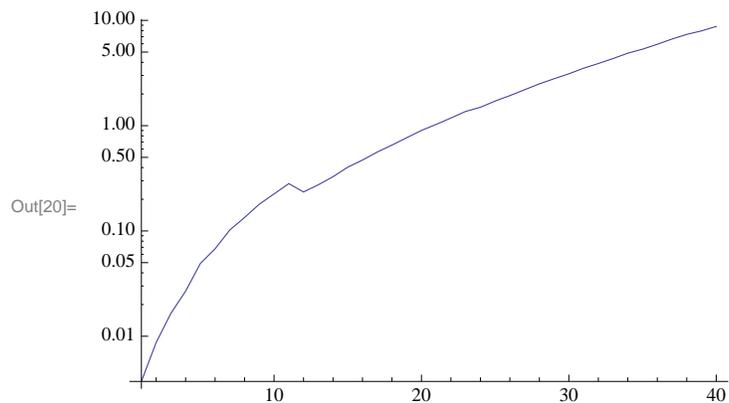

In[21]:= `TimeRatio40to20 = NNChain[40, -1][[4]] / NNChain[20, -1][[4]]`

Out[21]= `7.64344`

In[22]:= `NumberStatesRatio40to20 = 2^40 / 2^20`

Out[22]= `1 048 576`

# CplexA tutorial (*Mathematica* implementation)

*Jose M. G. Vilar* [1,2] *and Leonor Saiz* [3]

[1]*Biophysics Unit (CSIC-UPV/EHU) and Department of Biochemistry and Molecular Biology, University of the Basque Country, P.O. Box 644, 48080 Bilbao, Spain*

[2]*IKERBASQUE, Basque Foundation for Science, 48011 Bilbao, Spain*

[3]*Department of Biomedical Engineering, University of California, 451 East Health Sciences Drive, Davis, CA 95616, USA*

SUMMARY: CplexA is a software package (available for Mathematica, Python, and Matlab) to study the effects of macromolecular assembly on cellular processes. A main challenge for conventional computational methods to study macromolecular assembly is tackling the exponential increase of the number of configurational states with the number of components. CplexA uses functional programming to efficiently compute probabilities and average properties over such exponentially large number of states from the energetics of the interactions. The package is particularly suited to study gene expression at complex promoters controlled by multiple, local and distal, DNA binding sites for transcription factors. CplexA is freely available at http://sourceforge.net/projects/cplexa/.

REFERENCES:
1) The main reference for CplexA is Vilar, J.M.G. and Saiz, L. (2010) CplexA: a Mathematica package to study macromolecular-assembly control of gene expression, *Bioinformatics* 26, 2060-2061.
2) The biophysical background of the methods used by CplexA are described in Saiz, L. and Vilar, J.M.G. (2006) Stochastic dynamics of macromolecular-assembly networks, *Mol Syst Biol*, 2, 2006.0024.

## Running the tutorial

To run the tutorial the file CplexA.m must be in the Path, the default list of directories searched by *Mathematica* to find an external file. A possibility is to place CplexA.m in the same directory as the tutorial notebook.

```
SetDirectory@NotebookDirectory[];

<< CplexA`
```

The main function of CplexA is **AvConf[Γ, ΔG, S, RT]**= $(\sum \Gamma e^{-\Delta G/RT})/(\sum e^{-\Delta G/RT})$, which efficiently computes the thermodynamic average of $\Gamma$ for a system with free energy $\Delta G$ described by the set S of state variables. Each state variable can take the values 0 or 1 to represent whether a property is present or not. Both $\Gamma$ and $\Delta G$ are functions of S. RT is the thermal energy (gas constant, R, times the absolute temperature, T). The sums are performed over all possible combinations of values of the state variables.

Note that different units can be used for the thermal energy, for instance kT (AvConf[Γ, ΔG, S, kT]) or even numerical values explicitly (AvConf[Γ, ΔG, S, 0.6]). The three-argument function **AvConf[Γ, ΔG, S]** assumes the default thermal energy, which is set initially to RT. The default thermal energy can be changed with the function **SetTE[default_value]**, as for instance SetTE[kT], SetTE[0.6], or SetTE[RT].

CplexA also implements the function **AvConfF[Γ, ΔG, S, RT]**, which produces the same results as AvConf[Γ, ΔG, S, RT], but uses a slightly different algorithm to compute the sums. It is slower than AvConf for small systems but sometimes can be faster for large systems.

Using CplexA it is also possible to follow the traditional approach where a table provides the free energies and probabilities of all the possible configurations of the complex. The function **DGTable[ΔG, S, RT]** of CplexA constructs a table with the free energy and statistical weight (Boltzmann factor) of each state that has a non-zero probability.

## An activator

A transcription factor with concentration n1 binds a DNA site described by the state variable s1. The variable indicates whether (s1=1) or not (s1=0) the transcription factor is bound. S is the set of state variables describing the system (in this case, it consists only of s1); $\Delta$G is the free energy of the system; $\Delta$G° is the standard free energy of binding at 1M; and $\Gamma$ is the transcription rate, which is $\Gamma$max when the transcription factor is bound (s1=1) or zero when it is not (s1=0).

The example below computes the average transcription rate A$\Gamma$.

```
S := {s1};
ΔG := (ΔG° - RT Log[n1]) s1;
Γ := Γmax s1;

AΓ = AvConf[Γ, ΔG, S]
```

$$\frac{n1 \; \Gamma max}{e^{\frac{\Delta G°}{RT}} + n1}$$

Same example as before, but now the free energy is expressed in terms of the dissociation constant Kd (recall the equivalency Kd=$e^{\frac{\Delta G°}{RT}}$):

```
ΔG := - RT Log[n1 / Kd] s1;

AΓ = AvConf[Γ, ΔG, S, RT]
```

$$\frac{n1 \; \Gamma max}{Kd + n1}$$

## Two molecules binding to overlapping sites

Two molecules with concentrations n1 and n2 compete for binding to overlapping sites (binding of one excludes the binding of the other). The variables s1 and s2 indicate whether molecules 1 and 2 are bound, respectively. The expression below computes the probability that molecule 1 is bound, which is given by the average value of s1.

Note the term $\infty$ s1 s2 in the free energy. It implements the logical condition that two molecules cannot be bound simultaneously by assigning an infinite free energy to the corresponding state. To obtain the correct thermodynamic result, the symbol $\infty$ is considered to be finite until the end result, and then the limit to infinity is taken.

```
FullSimplify[AvConf[s1, -RT Log[n1 / K1] s1 - RT Log[n2 / K2] s2 + ∞ s1 s2, {s1, s2}, RT]]
```

$$\frac{K2 \; n1}{K2 \; n1 + K1 \; (K2 + n2)}$$

CplexA also provides the function DGTable[$\Delta$G, S, RT], which constructs a table with the free energy and statistical weight (Boltzmann factor) of each state that has a non-zero probability. For the previous example, it is used as follows:

```
DGTable[- RT Log[n1 / K1] s1 - RT Log[n2 / K2] s2 + ∞ s1 s2, {s1, s2}, RT]
```

| $\Delta G$ | $e^{-\frac{\Delta G}{RT}}$ | ( s1  s2 ) |
|---|---|---|
| 0 | 1 | ( 0  0 ) |
| $- RT Log\left[\frac{n2}{K2}\right]$ | $\frac{n2}{K2}$ | ( 0  1 ) |
| $- RT Log\left[\frac{n1}{K1}\right]$ | $\frac{n1}{K1}$ | ( 1  0 ) |

## Cooperative binding

Similar to the previous example, but now the binding of one does not exclude the other. The term e12 is the interaction energy between the molecules bound. The following expression computes the probability of having the two sites occupied simultaneously:

```
AvConf[s1 s2, - RT Log[n1 / K1] s1 - RT Log[n2 / K2] s2 + e12 s1 s2, {s1, s2}, RT]
```

$$\frac{n1\ n2}{n1\ n2 + e^{e12/RT}\ (K2\ n1 + K1\ (K2 + n2))}$$

The function DGTable[ΔG, S, RT] can be used here to construct the free energy and statistical weight table for cooperative binding:

```
DGTable[- RT Log[n1 / K1] s1 - RT Log[n2 / K2] s2 + e12 s1 s2, {s1, s2}, RT]
```

| $\Delta G$ | $e^{-\frac{\Delta G}{RT}}$ | ( s1  s2 ) |
|---|---|---|
| 0 | 1 | ( 0  0 ) |
| $- RT Log\left[\frac{n2}{K2}\right]$ | $\frac{n2}{K2}$ | ( 0  1 ) |
| $- RT Log\left[\frac{n1}{K1}\right]$ | $\frac{n1}{K1}$ | ( 1  0 ) |
| $e12 - RT Log\left[\frac{n1}{K1}\right] - RT Log\left[\frac{n2}{K2}\right]$ | $\frac{e^{\frac{e12}{RT}}\ n1\ n2}{K1\ K2}$ | ( 1  1 ) |

## *lac* Operon

The lac repressor is a protein with two DNA binding domains. It can bind to one site or, by looping the intervening DNA, to two distal sites simultaneously.

In addition to the domain-binding state variables to each site, s1 and s2, there is a looping state variable sL that indicates wether the DNA is looped (sL=1) or not (sL=0). Note that the each binding domain at sites 1 and 2 can be from two different lac repressors or from a single one when there is looping. The term ∞ (1 - s1 s2) sL means that looping can only happen with two domains bound, otherwise the free energy is infinite.
The following expression computes the repression level:

```
FullSimplify[1 / AvConf[(1 - s1), - RT Log[n / K1] s1 -
    RT Log[n / K2] s2 + (gL + RT Log[n] s1 s2) sL + ∞ (1 - s1 s2) sL, {s1, s2, sL}, RT]]
```

$$1 + \frac{n + \frac{e^{\frac{gL}{RT}}\ n}{K2 + n}}{K1}$$

Note that gL is the contribution to the free energy due to the conformational change of DNA (free energy of looping).

The following expression constructs the table for the free energies and statistical weights for the binding of the lac repressor to two DNA binding sites:

```
DGTable[-RT Log[n / K1] s1 - RT Log[n / K2] s2 +
  (gL + RT Log[n] s1 s2) sL + ∞ (1 - s1 s2) sL, {s1, s2, sL}, RT]
```

| $\Delta G$ | $e^{-\frac{\Delta G}{RT}}$ | ( s1 | s2 | sL ) |
|---|---|---|---|---|
| 0 | 1 | ( 0 | 0 | 0 ) |
| $- RT \, Log\left[\frac{n}{K2}\right]$ | $\frac{n}{K2}$ | ( 0 | 1 | 0 ) |
| $- RT \, Log\left[\frac{n}{K1}\right]$ | $\frac{n}{K1}$ | ( 1 | 0 | 0 ) |
| $- RT \, \left(Log\left[\frac{n}{K1}\right] + Log\left[\frac{n}{K2}\right]\right)$ | $\frac{n^2}{K1 \, K2}$ | ( 1 | 1 | 0 ) |
| $gL + RT \, Log[n] - RT \, Log\left[\frac{n}{K1}\right] - RT \, Log\left[\frac{n}{K2}\right]$ | $\frac{e^{-\frac{gL}{RT}} \, n}{K1 \, K2}$ | ( 1 | 1 | 1 ) |

## Phage $\lambda$

The CI protein of phage $\lambda$ can bind as a dimer to three sites on the right operator (described here by the state variables sOR1, sOR2, and sOR3 ) and to another three sites on the left operator (described by sOL1, sOL2, and sOL3). The state variable sL describes the looping state.

The free energy of the system $\Delta G\lambda$ takes into account interactions between CI dimers bound at neighboring sites and, when there is looping (sL=1), at different operators as well.

```
S := {sOL1, sOL2, sOL3, sOR1, sOR2, sOR3, sL}

ΔGλ := sOR1 (-12.7 - RT Log[n]) + sOR2 (-10.7 - RT Log[n]) + sOR3 (-10.2 - RT Log[n]) -
    3 sOR1 sOR2 - 3 sOR2 sOR3 + 3 sOR1 sOR2 sOR3 + sOL1 (-13.8 - RT Log[n]) +
    sOL2 (-12.1 - RT Log[n]) + sOL3 (-12.4 - RT Log[n]) - 2.5 sOL1 sOL2 - 2.5 sOL2 sOL3 +
    2.5 sOL1 sOL2 sOL3 + sL (21 - 21.2 sOL1 sOL2 sOR1 sOR2 - 3 sOL3 sOR3);
```

Transcription at the PRM promoter is given by Γrm, which leads to the average AΓrm. There is also another promoter, PR, with transcription given by Γr, which leads to the average AΓr.

```
Γrm := (1 - sOR3) (45 + ((240 - 45) sL + (460 - 45) (1 - sL)) sOR2);

AΓrm = AvConf[Γrm, ΔGλ, S];

AΓrm /. RT → 0.6
```

$\left(1.43575 \times 10^{-46} \left(3.13426 \times 10^{47} + 4.19648 \times 10^{57} \, n + 1.75428 \times 10^{68} \, n^2 + 8.39937 \times 10^{77} \, n^3 + 2.76788 \times 10^{88} \, n^4 + 2.42238 \times 10^{97} \, n^5\right)\right) / $
$\left(1. + 1.29007 \times 10^{10} \, n + 4.35928 \times 10^{20} \, n^2 + 1.15044 \times 10^{30} \, n^3 + 1.23153 \times 10^{40} \, n^4 + 1.08432 \times 10^{49} \, n^5 + 2.19947 \times 10^{58} \, n^6\right)$

```
Γr := (1 - sOR1) 1200;

AΓr = AvConf[Γr, ΔGλ, S];
```

```
AΓr /. RT → 0.6
```

$$\left(5.34231 \times 10^{-46} \left(2.24622 \times 10^{48} + 2.54781 \times 10^{58}\, n +\right.\right.$$
$$\left.9.10823 \times 10^{68}\, n^2 + 8.42144 \times 10^{77}\, n^3 + 2.41961 \times 10^{86}\, n^4 + 1.52361 \times 10^{95}\, n^5\right)\right) \Big/$$
$$\left(1. + 1.29007 \times 10^{10}\, n + 4.35928 \times 10^{20}\, n^2 + 1.15044 \times 10^{30}\, n^3 + 1.23153 \times 10^{40}\, n^4 +\right.$$
$$\left.1.08432 \times 10^{49}\, n^5 + 2.19947 \times 10^{58}\, n^6\right)$$

Note that in both cases it is possible to obtain an exact analytical expression for the effective transcription rates, AΓrm and AΓr, as a function of the CI dimer concentration. The expression NF below gives the CI dimer concentration as function of the normalized CI monomer concentration.

```
NF[nn_] := 9.38419 * 10^-14 + 7.0922 * 10^-10 nn - 1.15373 * 10^-11 Sqrt[0.0000661586 + 1. nn]
```

Plotted below is AΓrm (solid red line) as a function of the normalized CI monomer concentration. The filled circles correspond the experimentally measured activity of the PRM promoter (Dodd, I.B., Shearwin, K.E., Perkins, A.J., Burr, T., Hochschild, A.and Egan, J.B.(2004) Cooperativity in longrange gene regulation by the lambda CI repressor, Genes Dev, 18, 344-354).

```
PRMdata := {{0.058, 66.367}, {0.373, 189.358}, {0.658, 214.233}, {0.922, 188.897},
    {1.168, 173.696}, {1.385, 155.271}, {1.875, 133.160}, {2.281, 136.384}};
plotPRM := ListPlot[PRMdata, PlotMarkers → {Automatic, 17}, PlotRange → {{0, 2.5}, {0, 250}},
    ImageSize → 300, BaseStyle → {FontFamily → "Helvetica", FontSize → 12}];

Show[plotPRM, Plot[AΓrm /. {RT → 0.6, n → NF[nn]}, {nn, 0, 2.5}, PlotStyle → {Red, Thick}],
    AxesLabel → {"[CI]", "AΓrm"}]
```

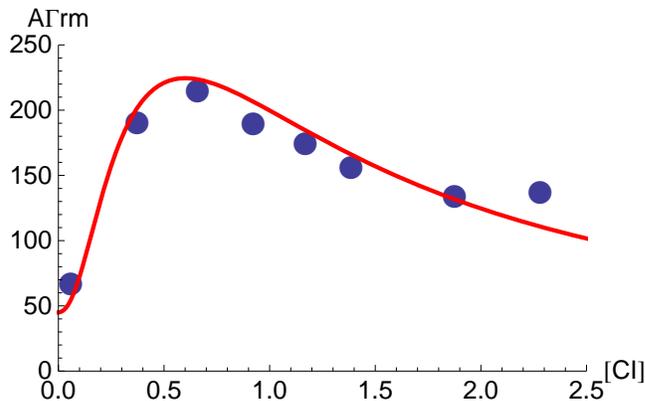

The expression below computes and plots the kinetics of the normalized CI concentration starting from [CI]=0 until it reaches its saturation value. It assumes that the production of CI is proportional to the transcriptional activity, AΓrm, and that the degradation rate of CI is equal to 1.

```
sol = NDSolve[{∂_t nn[t] == (AΓrm /. {RT → 0.6, n → NF[nn[t]]}) / 200 - nn[t], nn[0] == 0},
  nn, {t, 0, 5}]; Plot[Evaluate[nn[t] /. sol], {t, 0, 5}, PlotRange → All,
 PlotStyle → {Blue, Thick}, AxesLabel → {"Time", "[CI]"}, ImageSize → 300,
 BaseStyle → {FontFamily → "Helvetica", FontSize → 12}]
```

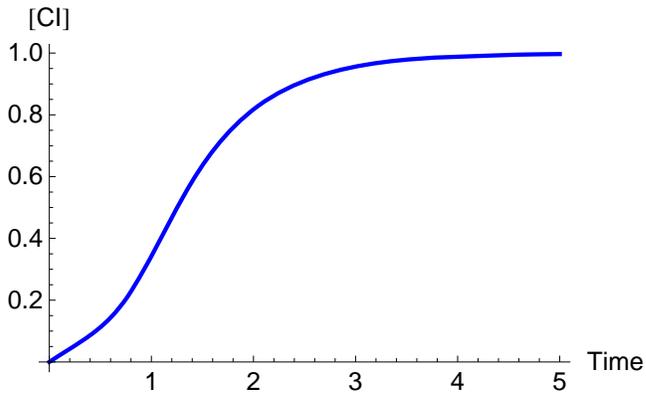

Plotted below is AΓr (solid red line) as a function of the normalized CI monomer concentration. The filled circles correspond the experimentally measured activity of the PR promoter ( Dodd, I.B., Shearwin, K.E., Perkins, A.J., Burr, T., Hochschild, A.and Egan, J.B.(2004) Cooperativity in longrange gene regulation by the lambda CI repressor, Genes Dev, 18, 344 - 354).

```
PRdata := {{0.063, 1057.627}, {0.373, 585.763}, {0.658, 204.746}, {0.921, 70.508},
    {1.171, 27.119}, {1.384, 18.983}, {1.875, 12.203}, {2.278, 9.492}};
plotPR := ListPlot[PRdata, PlotMarkers → {Automatic, 17}, PlotRange → {{0, 2.5}, {0, 1200}},
   ImageSize → 300, BaseStyle → {FontFamily → "Helvetica", FontSize → 12}];

Show[plotPR, Plot[AΓr /. {RT → 0.6, n → NF[nn]}, {nn, 0, 2.5},
   PlotStyle → {Red, Thick}, PlotRange → All], AxesLabel → {"[CI]", "AΓr"}]
```

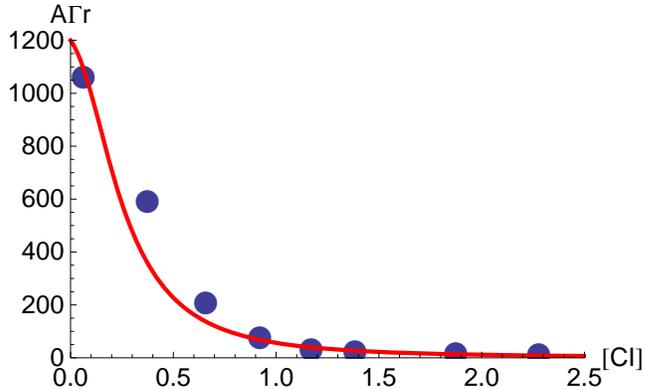

In this case,  CplexA is able to provide with a few lines of code results that would require writing tables with 128 entries, one for each binding state, with traditional thermodynamic approaches, as shown below.

```
DGTable[ΔGλ, S]
```

| $\Delta G$ | $e^{-\frac{\Delta G}{RT}}$ | ( sOL1 | sOL2 | sOL3 | sOR1 | sOR2 | sOR3 | sL ) |
|---|---|---|---|---|---|---|---|---|
| 0 | 1 | ( 0 | 0 | 0 | 0 | 0 | 0 | 0 ) |
| 21 | $e^{-21/RT}$ | ( 0 | 0 | 0 | 0 | 0 | 0 | 1 ) |
| $-10.2 - RT\,Log[n]$ | $e^{10.2/RT}\,n$ | ( 0 | 0 | 0 | 0 | 0 | 1 | 0 ) |
| $10.8 - RT\,Log[n]$ | $e^{-10.8/RT}\,n$ | ( 0 | 0 | 0 | 0 | 0 | 1 | 1 ) |
| $-10.7 - RT\,Log[n]$ | $e^{10.7/RT}\,n$ | ( 0 | 0 | 0 | 0 | 1 | 0 | 0 ) |
| $10.3 - RT\,Log[n]$ | $e^{-10.3/RT}\,n$ | ( 0 | 0 | 0 | 0 | 1 | 0 | 1 ) |
| $-23.9 - 2\,RT\,Log[n]$ | $e^{23.9/RT}\,n^2$ | ( 0 | 0 | 0 | 0 | 1 | 1 | 0 ) |
| $-2.9 - 2\,RT\,Log[n]$ | $e^{2.9/RT}\,n^2$ | ( 0 | 0 | 0 | 0 | 1 | 1 | 1 ) |
| $-12.7 - RT\,Log[n]$ | $e^{12.7/RT}\,n$ | ( 0 | 0 | 0 | 1 | 0 | 0 | 0 ) |
| $8.3 - RT\,Log[n]$ | $e^{-8.3/RT}\,n$ | ( 0 | 0 | 0 | 1 | 0 | 0 | 1 ) |
| $-22.9 - 2\,RT\,Log[n]$ | $e^{22.9/RT}\,n^2$ | ( 0 | 0 | 0 | 1 | 0 | 1 | 0 ) |
| $-1.9 - 2\,RT\,Log[n]$ | $e^{1.9/RT}\,n^2$ | ( 0 | 0 | 0 | 1 | 0 | 1 | 1 ) |
| $-26.4 - 2\,RT\,Log[n]$ | $e^{26.4/RT}\,n^2$ | ( 0 | 0 | 0 | 1 | 1 | 0 | 0 ) |
| $-5.4 - 2\,RT\,Log[n]$ | $e^{5.4/RT}\,n^2$ | ( 0 | 0 | 0 | 1 | 1 | 0 | 1 ) |
| $-36.6 - 3\,RT\,Log[n]$ | $e^{36.6/RT}\,n^3$ | ( 0 | 0 | 0 | 1 | 1 | 1 | 0 ) |
| $-15.6 - 3\,RT\,Log[n]$ | $e^{15.6/RT}\,n^3$ | ( 0 | 0 | 0 | 1 | 1 | 1 | 1 ) |
| $-12.4 - RT\,Log[n]$ | $e^{12.4/RT}\,n$ | ( 0 | 0 | 1 | 0 | 0 | 0 | 0 ) |
| $8.6 - RT\,Log[n]$ | $e^{-8.6/RT}\,n$ | ( 0 | 0 | 1 | 0 | 0 | 0 | 1 ) |
| $-22.6 - 2\,RT\,Log[n]$ | $e^{22.6/RT}\,n^2$ | ( 0 | 0 | 1 | 0 | 0 | 1 | 0 ) |
| $-4.6 - 2\,RT\,Log[n]$ | $e^{4.6/RT}\,n^2$ | ( 0 | 0 | 1 | 0 | 0 | 1 | 1 ) |
| $-23.1 - 2\,RT\,Log[n]$ | $e^{23.1/RT}\,n^2$ | ( 0 | 0 | 1 | 0 | 1 | 0 | 0 ) |
| $-2.1 - 2\,RT\,Log[n]$ | $e^{2.1/RT}\,n^2$ | ( 0 | 0 | 1 | 0 | 1 | 0 | 1 ) |
| $-36.3 - 3\,RT\,Log[n]$ | $e^{36.3/RT}\,n^3$ | ( 0 | 0 | 1 | 0 | 1 | 1 | 0 ) |
| $-18.3 - 3\,RT\,Log[n]$ | $e^{18.3/RT}\,n^3$ | ( 0 | 0 | 1 | 0 | 1 | 1 | 1 ) |
| $-25.1 - 2\,RT\,Log[n]$ | $e^{25.1/RT}\,n^2$ | ( 0 | 0 | 1 | 1 | 0 | 0 | 0 ) |
| $-4.1 - 2\,RT\,Log[n]$ | $e^{4.1/RT}\,n^2$ | ( 0 | 0 | 1 | 1 | 0 | 0 | 1 ) |
| $-35.3 - 3\,RT\,Log[n]$ | $e^{35.3/RT}\,n^3$ | ( 0 | 0 | 1 | 1 | 0 | 1 | 0 ) |
| $-17.3 - 3\,RT\,Log[n]$ | $e^{17.3/RT}\,n^3$ | ( 0 | 0 | 1 | 1 | 0 | 1 | 1 ) |
| $-38.8 - 3\,RT\,Log[n]$ | $e^{38.8/RT}\,n^3$ | ( 0 | 0 | 1 | 1 | 1 | 0 | 0 ) |
| $-17.8 - 3\,RT\,Log[n]$ | $e^{17.8/RT}\,n^3$ | ( 0 | 0 | 1 | 1 | 1 | 0 | 1 ) |
| $-49. - 4\,RT\,Log[n]$ | $e^{49./RT}\,n^4$ | ( 0 | 0 | 1 | 1 | 1 | 1 | 0 ) |
| $-31. - 4\,RT\,Log[n]$ | $e^{31./RT}\,n^4$ | ( 0 | 0 | 1 | 1 | 1 | 1 | 1 ) |
| $-12.1 - RT\,Log[n]$ | $e^{12.1/RT}\,n$ | ( 0 | 1 | 0 | 0 | 0 | 0 | 0 ) |
| $8.9 - RT\,Log[n]$ | $e^{-8.9/RT}\,n$ | ( 0 | 1 | 0 | 0 | 0 | 0 | 1 ) |
| $-22.3 - 2\,RT\,Log[n]$ | $e^{22.3/RT}\,n^2$ | ( 0 | 1 | 0 | 0 | 0 | 1 | 0 ) |
| $-1.3 - 2\,RT\,Log[n]$ | $e^{1.3/RT}\,n^2$ | ( 0 | 1 | 0 | 0 | 0 | 1 | 1 ) |
| $-22.8 - 2\,RT\,Log[n]$ | $e^{22.8/RT}\,n^2$ | ( 0 | 1 | 0 | 0 | 1 | 0 | 0 ) |

$-1.8 - 2\,RT\,Log[n]$   $e^{1.8/RT}\,n^2$   ( 0 1 0 0 1 0 1 )
$-36. - 3\,RT\,Log[n]$   $e^{36./RT}\,n^3$   ( 0 1 0 0 1 1 0 )
$-15. - 3\,RT\,Log[n]$   $e^{15./RT}\,n^3$   ( 0 1 0 0 1 1 1 )
$-24.8 - 2\,RT\,Log[n]$   $e^{24.8/RT}\,n^2$   ( 0 1 0 1 0 0 0 )
$-3.8 - 2\,RT\,Log[n]$   $e^{3.8/RT}\,n^2$   ( 0 1 0 1 0 0 1 )
$-35. - 3\,RT\,Log[n]$   $e^{35./RT}\,n^3$   ( 0 1 0 1 0 1 0 )
$-14. - 3\,RT\,Log[n]$   $e^{14./RT}\,n^3$   ( 0 1 0 1 0 1 1 )
$-38.5 - 3\,RT\,Log[n]$   $e^{38.5/RT}\,n^3$   ( 0 1 0 1 1 0 0 )
$-17.5 - 3\,RT\,Log[n]$   $e^{17.5/RT}\,n^3$   ( 0 1 0 1 1 0 1 )
$-48.7 - 4\,RT\,Log[n]$   $e^{48.7/RT}\,n^4$   ( 0 1 0 1 1 1 0 )
$-27.7 - 4\,RT\,Log[n]$   $e^{27.7/RT}\,n^4$   ( 0 1 0 1 1 1 1 )
$-27. - 2\,RT\,Log[n]$   $e^{27./RT}\,n^2$   ( 0 1 1 0 0 0 0 )
$-6. - 2\,RT\,Log[n]$   $e^{6./RT}\,n^2$   ( 0 1 1 0 0 0 1 )
$-37.2 - 3\,RT\,Log[n]$   $e^{37.2/RT}\,n^3$   ( 0 1 1 0 0 1 0 )
$-19.2 - 3\,RT\,Log[n]$   $e^{19.2/RT}\,n^3$   ( 0 1 1 0 0 1 1 )
$-37.7 - 3\,RT\,Log[n]$   $e^{37.7/RT}\,n^3$   ( 0 1 1 0 1 0 0 )
$-16.7 - 3\,RT\,Log[n]$   $e^{16.7/RT}\,n^3$   ( 0 1 1 0 1 0 1 )
$-50.9 - 4\,RT\,Log[n]$   $e^{50.9/RT}\,n^4$   ( 0 1 1 0 1 1 0 )
$-32.9 - 4\,RT\,Log[n]$   $e^{32.9/RT}\,n^4$   ( 0 1 1 0 1 1 1 )
$-39.7 - 3\,RT\,Log[n]$   $e^{39.7/RT}\,n^3$   ( 0 1 1 1 0 0 0 )
$-18.7 - 3\,RT\,Log[n]$   $e^{18.7/RT}\,n^3$   ( 0 1 1 1 0 0 1 )
$-49.9 - 4\,RT\,Log[n]$   $e^{49.9/RT}\,n^4$   ( 0 1 1 1 0 1 0 )
$-31.9 - 4\,RT\,Log[n]$   $e^{31.9/RT}\,n^4$   ( 0 1 1 1 0 1 1 )
$-53.4 - 4\,RT\,Log[n]$   $e^{53.4/RT}\,n^4$   ( 0 1 1 1 1 0 0 )
$-32.4 - 4\,RT\,Log[n]$   $e^{32.4/RT}\,n^4$   ( 0 1 1 1 1 0 1 )
$-63.6 - 5\,RT\,Log[n]$   $e^{63.6/RT}\,n^5$   ( 0 1 1 1 1 1 0 )
$-45.6 - 5\,RT\,Log[n]$   $e^{45.6/RT}\,n^5$   ( 0 1 1 1 1 1 1 )
$-13.8 - RT\,Log[n]$   $e^{13.8/RT}\,n$   ( 1 0 0 0 0 0 0 )
$7.2 - RT\,Log[n]$   $e^{-7.2/RT}\,n$   ( 1 0 0 0 0 0 1 )
$-24. - 2\,RT\,Log[n]$   $e^{24./RT}\,n^2$   ( 1 0 0 0 0 1 0 )
$-3. - 2\,RT\,Log[n]$   $e^{3./RT}\,n^2$   ( 1 0 0 0 0 1 1 )
$-24.5 - 2\,RT\,Log[n]$   $e^{24.5/RT}\,n^2$   ( 1 0 0 0 1 0 0 )
$-3.5 - 2\,RT\,Log[n]$   $e^{3.5/RT}\,n^2$   ( 1 0 0 0 1 0 1 )
$-37.7 - 3\,RT\,Log[n]$   $e^{37.7/RT}\,n^3$   ( 1 0 0 0 1 1 0 )
$-16.7 - 3\,RT\,Log[n]$   $e^{16.7/RT}\,n^3$   ( 1 0 0 0 1 1 1 )
$-26.5 - 2\,RT\,Log[n]$   $e^{26.5/RT}\,n^2$   ( 1 0 0 1 0 0 0 )
$-5.5 - 2\,RT\,Log[n]$   $e^{5.5/RT}\,n^2$   ( 1 0 0 1 0 0 1 )
$-36.7 - 3\,RT\,Log[n]$   $e^{36.7/RT}\,n^3$   ( 1 0 0 1 0 1 0 )

$-15.7 - 3\,RT\,Log[n]$    $e^{15.7/RT}\,n^3$    ( 1 0 0 1 0 1 1 )

$-40.2 - 3\,RT\,Log[n]$    $e^{40.2/RT}\,n^3$    ( 1 0 0 1 1 0 0 )

$-19.2 - 3\,RT\,Log[n]$    $e^{19.2/RT}\,n^3$    ( 1 0 0 1 1 0 1 )

$-50.4 - 4\,RT\,Log[n]$    $e^{50.4/RT}\,n^4$    ( 1 0 0 1 1 1 0 )

$-29.4 - 4\,RT\,Log[n]$    $e^{29.4/RT}\,n^4$    ( 1 0 0 1 1 1 1 )

$-26.2 - 2\,RT\,Log[n]$    $e^{26.2/RT}\,n^2$    ( 1 0 1 0 0 0 0 )

$-5.2 - 2\,RT\,Log[n]$    $e^{5.2/RT}\,n^2$    ( 1 0 1 0 0 0 1 )

$-36.4 - 3\,RT\,Log[n]$    $e^{36.4/RT}\,n^3$    ( 1 0 1 0 0 1 0 )

$-18.4 - 3\,RT\,Log[n]$    $e^{18.4/RT}\,n^3$    ( 1 0 1 0 0 1 1 )

$-36.9 - 3\,RT\,Log[n]$    $e^{36.9/RT}\,n^3$    ( 1 0 1 0 1 0 0 )

$-15.9 - 3\,RT\,Log[n]$    $e^{15.9/RT}\,n^3$    ( 1 0 1 0 1 0 1 )

$-50.1 - 4\,RT\,Log[n]$    $e^{50.1/RT}\,n^4$    ( 1 0 1 0 1 1 0 )

$-32.1 - 4\,RT\,Log[n]$    $e^{32.1/RT}\,n^4$    ( 1 0 1 0 1 1 1 )

$-38.9 - 3\,RT\,Log[n]$    $e^{38.9/RT}\,n^3$    ( 1 0 1 1 0 0 0 )

$-17.9 - 3\,RT\,Log[n]$    $e^{17.9/RT}\,n^3$    ( 1 0 1 1 0 0 1 )

$-49.1 - 4\,RT\,Log[n]$    $e^{49.1/RT}\,n^4$    ( 1 0 1 1 0 1 0 )

$-31.1 - 4\,RT\,Log[n]$    $e^{31.1/RT}\,n^4$    ( 1 0 1 1 0 1 1 )

$-52.6 - 4\,RT\,Log[n]$    $e^{52.6/RT}\,n^4$    ( 1 0 1 1 1 0 0 )

$-31.6 - 4\,RT\,Log[n]$    $e^{31.6/RT}\,n^4$    ( 1 0 1 1 1 0 1 )

$-62.8 - 5\,RT\,Log[n]$    $e^{62.8/RT}\,n^5$    ( 1 0 1 1 1 1 0 )

$-44.8 - 5\,RT\,Log[n]$    $e^{44.8/RT}\,n^5$    ( 1 0 1 1 1 1 1 )

$-28.4 - 2\,RT\,Log[n]$    $e^{28.4/RT}\,n^2$    ( 1 1 0 0 0 0 0 )

$-7.4 - 2\,RT\,Log[n]$    $e^{7.4/RT}\,n^2$    ( 1 1 0 0 0 0 1 )

$-38.6 - 3\,RT\,Log[n]$    $e^{38.6/RT}\,n^3$    ( 1 1 0 0 0 1 0 )

$-17.6 - 3\,RT\,Log[n]$    $e^{17.6/RT}\,n^3$    ( 1 1 0 0 0 1 1 )

$-39.1 - 3\,RT\,Log[n]$    $e^{39.1/RT}\,n^3$    ( 1 1 0 0 1 0 0 )

$-18.1 - 3\,RT\,Log[n]$    $e^{18.1/RT}\,n^3$    ( 1 1 0 0 1 0 1 )

$-52.3 - 4\,RT\,Log[n]$    $e^{52.3/RT}\,n^4$    ( 1 1 0 0 1 1 0 )

$-31.3 - 4\,RT\,Log[n]$    $e^{31.3/RT}\,n^4$    ( 1 1 0 0 1 1 1 )

$-41.1 - 3\,RT\,Log[n]$    $e^{41.1/RT}\,n^3$    ( 1 1 0 1 0 0 0 )

$-20.1 - 3\,RT\,Log[n]$    $e^{20.1/RT}\,n^3$    ( 1 1 0 1 0 0 1 )

$-51.3 - 4\,RT\,Log[n]$    $e^{51.3/RT}\,n^4$    ( 1 1 0 1 0 1 0 )

$-30.3 - 4\,RT\,Log[n]$    $e^{30.3/RT}\,n^4$    ( 1 1 0 1 0 1 1 )

$-54.8 - 4\,RT\,Log[n]$    $e^{54.8/RT}\,n^4$    ( 1 1 0 1 1 0 0 )

$-55. - 4\,RT\,Log[n]$    $e^{55./RT}\,n^4$    ( 1 1 0 1 1 0 1 )

$-65. - 5\,RT\,Log[n]$    $e^{65./RT}\,n^5$    ( 1 1 0 1 1 1 0 )

$-65.2 - 5\,RT\,Log[n]$    $e^{65.2/RT}\,n^5$    ( 1 1 0 1 1 1 1 )

$-40.8 - 3\,RT\,Log[n]$    $e^{40.8/RT}\,n^3$    ( 1 1 1 0 0 0 0 )

$-19.8 - 3\,RT\,Log[n]$   $e^{19.8/RT}\,n^3$   $(1\ 1\ 1\ 0\ 0\ 0\ 1)$

$-51. - 4\,RT\,Log[n]$   $e^{51./RT}\,n^4$   $(1\ 1\ 1\ 0\ 0\ 1\ 0)$

$-33. - 4\,RT\,Log[n]$   $e^{33./RT}\,n^4$   $(1\ 1\ 1\ 0\ 0\ 1\ 1)$

$-51.5 - 4\,RT\,Log[n]$   $e^{51.5/RT}\,n^4$   $(1\ 1\ 1\ 0\ 1\ 0\ 0)$

$-30.5 - 4\,RT\,Log[n]$   $e^{30.5/RT}\,n^4$   $(1\ 1\ 1\ 0\ 1\ 0\ 1)$

$-64.7 - 5\,RT\,Log[n]$   $e^{64.7/RT}\,n^5$   $(1\ 1\ 1\ 0\ 1\ 1\ 0)$

$-46.7 - 5\,RT\,Log[n]$   $e^{46.7/RT}\,n^5$   $(1\ 1\ 1\ 0\ 1\ 1\ 1)$

$-53.5 - 4\,RT\,Log[n]$   $e^{53.5/RT}\,n^4$   $(1\ 1\ 1\ 1\ 0\ 0\ 0)$

$-32.5 - 4\,RT\,Log[n]$   $e^{32.5/RT}\,n^4$   $(1\ 1\ 1\ 1\ 0\ 0\ 1)$

$-63.7 - 5\,RT\,Log[n]$   $e^{63.7/RT}\,n^5$   $(1\ 1\ 1\ 1\ 0\ 1\ 0)$

$-45.7 - 5\,RT\,Log[n]$   $e^{45.7/RT}\,n^5$   $(1\ 1\ 1\ 1\ 0\ 1\ 1)$

$-67.2 - 5\,RT\,Log[n]$   $e^{67.2/RT}\,n^5$   $(1\ 1\ 1\ 1\ 1\ 0\ 0)$

$-67.4 - 5\,RT\,Log[n]$   $e^{67.4/RT}\,n^5$   $(1\ 1\ 1\ 1\ 1\ 0\ 1)$

$-77.4 - 6\,RT\,Log[n]$   $e^{77.4/RT}\,n^6$   $(1\ 1\ 1\ 1\ 1\ 1\ 0)$

$-80.6 - 6\,RT\,Log[n]$   $e^{80.6/RT}\,n^6$   $(1\ 1\ 1\ 1\ 1\ 1\ 1)$

# CplexA tutorial (Matlab implementation)

## Table of Contents

*Jose M. G. Vilar 1,2 and Leonor Saiz 3*

*1. Biophysics Unit (CSIC-UPV/EHU) and Department of Biochemistry and Molecular Biology, University of the Basque Country, P.O. Box 644, 48080 Bilbao, Spain*

*2. IKERBASQUE, Basque Foundation for Science, 48011 Bilbao, Spain*

*3. Department of Biomedical Engineering, University of California, 451 East Health Sciences Drive, Davis, CA 95616, USA*

SUMMARY:

CplexA is a software package (available for Mathematica, Python, and Matlab) to study the effects of macromolecular assembly on cellular processes. A main challenge for conventional computational methods to study macromolecular assembly is tackling the exponential increase of the number of configurational states with the number of components. CplexA uses functional programming to efficiently compute probabilities and average properties over such exponentially large number of states from the energetics of the interactions. The package is particularly suited to study gene expression at complex promoters controlled by multiple, local and distal, DNA binding sites for transcription factors. CplexA is freely available at http://sourceforge.net/projects/cplexa/.

REFERENCES:

1. The main reference for CplexA is Vilar, J.M.G. and Saiz, L. (2010) CplexA: a Mathematica package to study macromolecular-assembly control of gene expression, *Bioinformatics* 26, 2060-2061.

2. The biophysical background of the methods used by CplexA are described in Saiz, L. and Vilar, J.M.G. (2006) Stochastic dynamics of macromolecular-assembly networks, *Mol Syst Biol*, 2, 2006.0024.

# Running the tutorial

To run the python tutorial, the file AveConf.m must be in the Path, the default list of directories searched by Matlab to find an external file. CplexA uses the *Symbolic Math Toolbox*.

The main function of CplexA is **AvConf(F, $\Delta$ G, S, RT)**$= \left( \sum F e^{-\Delta G/RT} \right) / \left( \sum e^{-\Delta G/RT} \right)$, which efficiently computes the thermodynamic average of F for a system with free energy $\Delta$ G described by the set S of state variables. The function returns a symbolic expression and the terms F, $\Delta$ G, S, and RT can

---

be symbolic or string expressions. Each state variable can take the values 0 or 1 to represent whether a property is present or not. Both F and $\Delta$ G are functions of S. RT is the thermal energy (gas constant, R, times the absolute temperature, T). The sums are performed over all possible combinations of values of the state variables.

Note that different units can be used for the thermal energy, for instance kT (AvConf(F, $\Delta$ G, S, kT)) or even numerical values explicitly (AvConf(F, $\Delta$ G, S, 0.6)). The three-argument function **AvConf(F, $\Delta$ G, S)** assumes RT as a default thermal energy.

# An activator

A transcription factor with concentration n1 binds a DNA site described by the state variable s1. The variable indicates whether (s1=1) or not (s1=0) the transcription factor is bound. S is the set of state variables describing the system (in this case, it consists only of s1); DG is the free energy of the system; DG0 is the standard free energy of binding at 1M; and F is the transcription rate, which is Fmax when the transcription factor is bound (s1=1) or zero when it is not (s1=0).

The example below computes the average transcription rate AF.

```
S='s1';
DG='(DG0-RT*log(n1))*s1';
F='Fmax*s1';
AF=AveConf(F,DG,S);

disp(simplify(AF))

        (Fmax*n1)/(n1 + exp(DG0/RT))
```

Same example as before, but now the free energy is expressed in terms of the dissociation constant Kd (recall the equivalency $Kd = E^{D(0/RT)}$:

```
DG='-RT*log(n1/Kd)*s1';
AF=AveConf(F,DG,S);

disp(AF)

        (Fmax*n1)/(Kd + n1)
```

# Two molecules binding to overlapping sites

Two molecules with concentrations n1 and n2 compete for binding to overlapping sites (binding of one excludes the binding of the other). The variables s1 and s2 indicate whether molecules 1 and 2 are bound, respectively. The expression below computes the probability that molecule 1 is bound, which is given by the average value of s1.

Note the term inf s1 s2 in the free energy. It implements the logical condition that two molecules cannot be bound simultaneously by assigning an infinite free energy to the corresponding state. To obtain the correct thermodynamic result, the symbol inf is considered to be finite until the end result, and then the limit to infinity is taken.

```
AF=AveConf('s1','-RT*log(n1/K1)*s1-RT*log(n2/K2)*s2+inf*RT*s1*s2','s1 s2');
```

```
disp(simplify(AF))

        (K2*n1)/(K2*n1 + K1*(K2 + n2))
```

# Cooperative binding

Similar to the previous example, but now the binding of one does not exclude the other. The term e12 is the interaction energy between the molecules bound. The following expression computes the probability of having the two sites occupied simultaneously:

```
AF=AveConf('s1*s2','-RT*log(n1/K1)*s1-RT*log(n2/K2)*s2+e12*s1*s2','s1 s2');

disp(simplify(AF))

        (n1*n2)/(exp(e12/RT)*(K1*n2 + K2*n1 + K1*K2 + (n1*n2)/exp(e12/RT)))
```

# *lac* Operon

The lac repressor is a protein with two DNA binding domains. It can bind to one site or, by looping the intervening DNA, to two distal sites simultaneously.

In addition to the domain-binding state variables to each site, s1 and s2, there is a looping state variable sL that indicates wether the DNA is looped (sL=1) or not (sL=0). Note that the each binding domain at sites 1 and 2 can be from two different lac repressors or from a single one when there is looping. The term inf (1 - s1 s2) sL means that looping can only happen with two domains bound, otherwise the free energy is infinite. The following expression computes the repression level:

```
AF=AveConf('(1-s1)',['-RT*log(n/K1)*s1-RT*log(n/K2)*s2'...
          ,'+(gL+RT*log(n)*s1*s2)*sL+inf*RT*(1-s1*s2)*sL'],'s1 s2 sL');

disp(1+simplify(1/AF-1))
disp(1+simplify(1/subs(AF,'gL','-RT*log(cL)')-1))

        (n*(K2 + n + 1/exp(gL/RT)))/(K1*(K2 + n)) + 1

        (n*(K2 + cL + n))/(K1*(K2 + n)) + 1
```

Note that gL is the contribution to the free energy due to the conformational change of DNA (free energy of looping).

The repression level can also be expressed in terms of the *local concentration* cL, defined as $c_L = e^{-g_L/RT}$.

# Phage $\lambda$

The CI protein of phage $\lambda$ can bind as a dimer to three sites on the right operator (described here by the state variables sOR1, sOR2, and sOR3 ) and to another three sites on the left operator (described by sOL1, sOL2, and sOL3). The state variable sL describes the looping state.

The free energy of the system DG takes into account interactions between CI dimers bound at neighboring sites and, when there is looping (sL=1), at different operators as well.

```
S='sOL1 sOL2 sOL3 sOR1 sOR2 sOR3 sL';
DG=['-2.5*sOL1*sOL2 - 2.5*sOL2*sOL3 + 2.5*sOL1*sOL2*sOL3'...
    ,'-3*sOR1*sOR2 - 3*sOR2*sOR3 + 3*sOR1*sOR2*sOR3'...
    ,'+ sL*(21 - 21.2*sOL1*sOL3*sOR1*sOR2 - 3*sOL3*sOR3)'...
    ,'+ sOL1*(-13.8 - RT*log(n)) + sOR1*(-12.7 - RT*log(n))'...
    ,'+ sOL3*(-12.4 - RT*log(n)) + sOL2*(-12.1 - RT*log(n))'...
    ,'+ sOR2*(-10.7 - RT*log(n)) + sOR3*(-10.2 - RT*log(n))'];
```

Transcription at the PRM promoter is given by F_rm, which leads to the average AF_rm. There is also another promoter, PR, with transcription given by F_r, which leads to the average AF_r.

```
F_rm='(45 + (415*(1 - sL) + 195*sL)*sOR2)*(1 - sOR3)';
AF_rm=AveConf(F_rm,DG,S);
AF_rm_n=simplify(subs(AF_rm,'RT',0.6));

disp(vpa(AF_rm_n,5))
```

$$(3.4779e51*n^5 + 3.974e42*n^4 + 1.2059e32*n^3 + 2.5187e22*n^2 + 6.0251e11*$$

```
F_r='1200*(1 - sOR1)';
AF_r=AveConf(F_r,DG,S);
AF_r_n=simplify(subs(AF_r,'RT',0.6));

disp(vpa(AF_r_n,5))
```

$$(8.1396e49*n^5 + 1.2926e41*n^4 + 4.499e32*n^3 + 4.8659e23*n^2 + 1.3611e13*$$

Note that in both cases it is possible to obtain an exact analytical expression for the effective transcription rates, AF_rm and AF_r, as a function of the CI dimer concentration. The expression nvals below gives the CI dimer concentration from the normalized CI monomer concentration nnvals.

```
nnvals=(0:0.025:2.5);
nvals=9.38419e-14 + 7.0922e-10*nnvals - 1.15373e-11*sqrt(0.0000661586 + 1.*nnvals)
```

Plotted below is AF_rm (solid red line) as a function of the normalized CI monomer concentration. The circles correspond the experimentally measured activity of the PRM promoter (Dodd, I.B., Shearwin, K.E., Perkins, A.J., Burr, T., Hochschild, A.and Egan, J.B.(2004) Cooperativity in longrange gene regulation by the lambda CI repressor, Genes Dev, 18, 344-354).

```
PRMdata = [0.058, 66.367; 0.373, 189.358; 0.658, 214.233; 0.922, 188.897;...
           1.168, 173.696; 1.385, 155.271; 1.875, 133.160; 2.281, 136.384];
ff=matlabFunction(AF_rm_n);
plot(nnvals,ff(nvals),'r--',PRMdata(:,1), PRMdata(:,2),'ob');
xlabel('[CI]','fontsize',14); ylabel('AF_{rm}','fontsize',14);
```

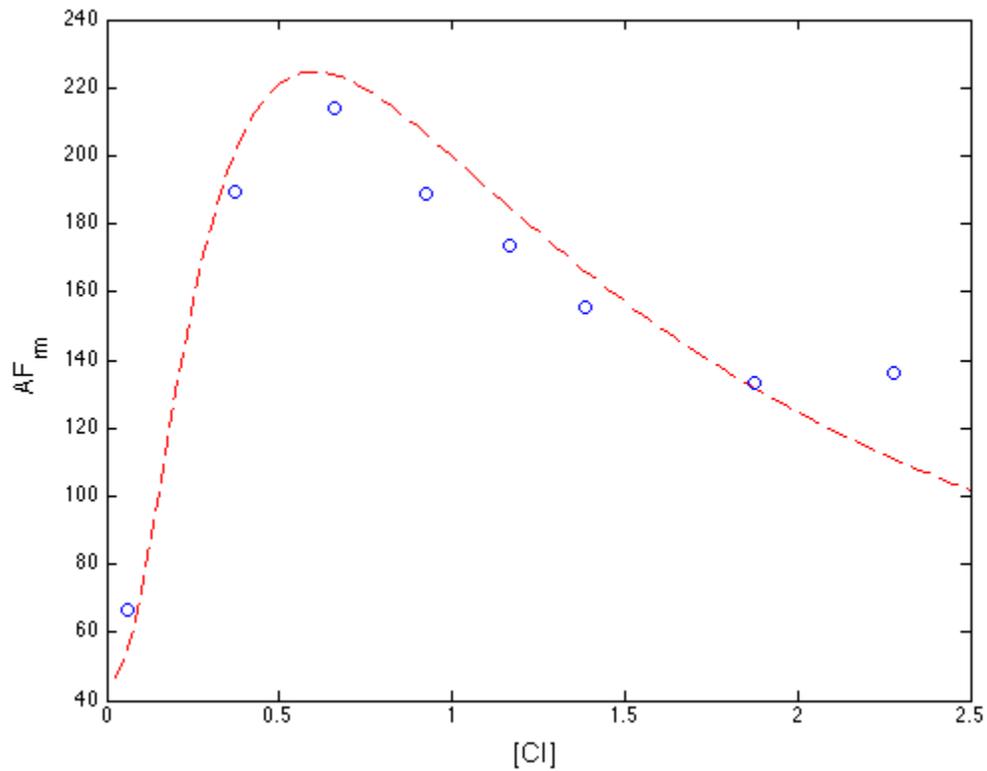

Plotted below is AF_r (solid red line) as a function of the normalized CI monomer concentration. The circles correspond the experimentally measured activity of the PR promoter ( Dodd, I.B., Shearwin, K.E., Perkins, A.J., Burr, T., Hochschild, A.and Egan, J.B.(2004) Cooperativity in longrange gene regulation by the lambda CI repressor, Genes Dev, 18, 344 - 354).

```
PRdata = [0.063, 1057.627; 0.373, 585.763; 0.658, 204.746; 0.921, 70.508;...
          1.171, 27.119; 1.384, 18.983; 1.875, 12.203; 2.278, 9.492];
ff=matlabFunction(AF_r_n);
plot(nnvals,ff(nvals),'r--',PRdata(:,1), PRdata(:,2),'ob');
xlabel('[CI]','fontsize',14); ylabel('AF_{r}','fontsize',14);
```

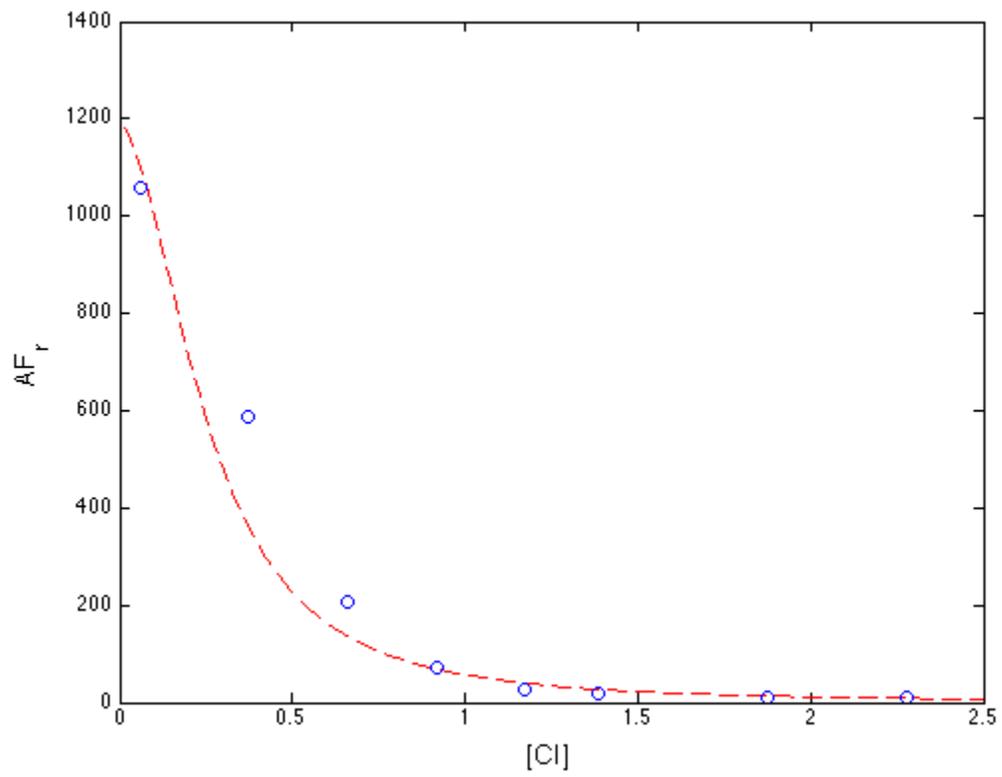

*Published with MATLAB® 7.13*

# CplexA tutorial (Python implementation)

*Jose M. G. Vilar[1,2] and Leonor Saiz[3]*

[1] *Biophysics Unit (CSIC-UPV/EHU) and Department of Biochemistry and Molecular Biology, University of the Basque Country, P.O. Box 644, 48080 Bilbao, Spain*

[2] *IKERBASQUE, Basque Foundation for Science, 48011 Bilbao, Spain*

[3] *Department of Biomedical Engineering, University of California, 451 East Health Sciences Drive, Davis, CA 95616, USA*

SUMMARY:
CplexA is a software package (available for Mathematica, Python, and Matlab) to study the effects of macromolecular assembly on cellular processes. A main challenge for conventional computational methods to study macromolecular assembly is tackling the exponential increase of the number of configurational states with the number of components. CplexA uses functional programming to efficiently compute probabilities and average properties over such exponentially large number of states from the energetics of the interactions. The package is particularly suited to study gene expression at complex promoters controlled by multiple, local and distal, DNA binding sites for transcription factors. CplexA is freely available at http://sourceforge.net/projects/cplexa/.

REFERENCES:
1. The main reference for CplexA is Vilar, J.M.G. and Saiz, L. (2010) CplexA: a Mathematica package to study macromolecular-assembly control of gene expression, *Bioinformatics* 26, 2060-2061.
2. The biophysical background of the methods used by CplexA are described in Saiz, L. and Vilar, J.M.G. (2006) Stochastic dynamics of macromolecular-assembly networks, *Mol Syst Biol*, 2, 2006.0024.

## Running the tutorial

To run the python tutorial, the file CplexA.py must be in the Path, the default list of directories searched by Python to find an external file. A possibility is to place CplexA.py in the same directory as the tutorial file. CplexA uses *SymPy*, a Python library for symbolic mathematics.

```
In [1]:  from pylab import *
         from numpy import *
         from sympy import *
         from CplexA import *
```

The main function of CplexA is **AvConf(F, $\Delta$G, S, RT)**= $\left(\sum F e^{-\Delta G/RT}\right)/\left(\sum e^{-\Delta G/RT}\right)$, which efficiently computes the thermodynamic average of F for a system with free energy $\Delta$G described by the set S of state variables. The function returns a SymPy expression and the terms F, $\Delta$G, S, and RT can be SymPy or string expressions. Each state variable can take the values 0 or 1 to represent whether a property is present or not. Both F and $\Delta$G are functions of S. RT is the thermal energy (gas constant, R, times the absolute temperature, T). The sums are performed over all possible combinations of values of the state variables.

Note that different units can be used for the thermal energy, for instance kT (AvConf(F, $\Delta$G, S, kT)) or even numerical values explicitly (AvConf(F, $\Delta$G, S, 0.6)). The three-argument function **AvConf(F, $\Delta$G, S)** assumes RT as a default thermal energy.

## An activator

A transcription factor with concentration n1 binds a DNA site described by the state variable s1. The variable indicates whether (s1=1) or not (s1=0) the transcription factor is bound. S is the set of state variables describing the system (in this case, it consists only of s1); DG is the free energy of the system; DG0 is the standard free energy of binding at 1M; and F is the transcription rate, which is Fmax when the transcription factor is bound (s1=1) or zero when it is not (s1=0).

The example below computes the average transcription rate AF.

```
In [2]:  S='s1'
         DG='(DG0-RT*ln(n1))*s1'
         F='Fmax*s1'
```

```
AF=AveConf(F,DG,S)

print simplify(AF)

Fmax*n1/(n1 + exp(DG0/RT))
```

Same example as before, but now the free energy is expressed in terms of the dissociation constant Kd (recall the equivalency $K_d = e^{\Delta G^\circ / RT}$):

```
In [3]:  DG='-RT*ln(n1/Kd)*s1'
         AF=AveConf(F,DG,S)

         print simplify(AF)

         Fmax*n1/(Kd + n1)
```

## Two molecules binding to overlapping sites

Two molecules with concentrations n1 and n2 compete for binding to overlapping sites (binding of one excludes the binding of the other). The variables s1 and s2 indicate whether molecules 1 and 2 are bound, respectively. The expression below computes the probability that molecule 1 is bound, which is given by the average value of s1.

Note the term inf s1 s2 in the free energy. It implements the logical condition that two molecules cannot be bound simultaneously by assigning an infinite free energy to the corresponding state. To obtain the correct thermodynamic result, the symbol "inf" is considered to be finite until the end result, and then the limit to infinity is taken.

```
In [4]:  AF=AveConf('s1','-RT*ln(n1/K1)*s1-RT*ln(n2/K2)*s2+inf*RT*s1*s2','s1 s2')

         print simplify(AF)

         K2*n1/(K1*K2 + K1*n2 + K2*n1)
```

## Cooperative binding

Similar to the previous example, but now the binding of one does not exclude the other. The term e12 is the interaction energy between the molecules bound. The following expression computes the probability of having the two sites occupied simultaneously:

```
In [5]:  AF=AveConf('s1*s2','-RT*ln(n1/K1)*s1-RT*ln(n2/K2)*s2+e12*s1*s2','s1 s2')

         print simplify(AF)

         n1*n2/(K1*K2*exp(e12/RT) + K1*n2*exp(e12/RT) + K2*n1*exp(e12/RT) + n1*n2)
```

## *lac* Operon

The lac repressor is a protein with two DNA binding domains. It can bind to one site or, by looping the intervening DNA, to two distal sites simultaneously.

In addition to the domain-binding state variables to each site, s1 and s2, there is a looping state variable sL that indicates wether the DNA is looped (sL=1) or not (sL=0). Note that the each binding domain at sites 1 and 2 can be from two different lac repressors or from a single one when there is looping. The term inf (1 - s1 s2) sL means that looping can only happen with two domains bound, otherwise the free energy is infinite. The following expression computes the repression level:

```
In [6]:  AF=AveConf('(1-s1)','-RT*ln(n/K1)*s1-RT*ln(n/K2)*s2\
                     +(gL+RT*ln(n)*s1*s2)*sL+inf*RT*(1-s1*s2)*sL','s1 s2 sL')
```

```
print 1+simplify(1/AF-1)

1 + n*(K2*exp(gL/RT) + n*exp(gL/RT) + 1)*exp(-gL/RT)/(K1*(K2 + n))
```

The quantity gL is the contribution to the free energy due to the conformational change of DNA (free energy of looping).

The repression level can also be expressed in terms of the *local concentration* cL, defined as $c_L = e^{-g_L/RT}$:

```
In [7]: print 1+simplify(subs(1/AF,'gL','-RT*log(cL)')-1)

1 + n*(K2 + cL + n)/(K1*(K2 + n))
```

# Phage $\lambda$

The CI protein of phage $\lambda$ can bind as a dimer to three sites on the right operator (described here by the state variables sOR1, sOR2, and sOR3 ) and to another three sites on the left operator (described by sOL1, sOL2, and sOL3). The state variable sL describes the looping state.

The free energy of the system DG takes into account interactions between CI dimers bound at neighboring sites and, when there is looping (sL=1), at different operators as well.

```
In [8]: S='sOL1 sOL2 sOL3 sOR1 sOR2 sOR3 sL'
        DG='-2.5*sOL1*sOL2 - 2.5*sOL2*sOL3 + 2.5*sOL1*sOL2*sOL3\
         - 3*sOR1*sOR2 - 3*sOR2*sOR3 + 3*sOR1*sOR2*sOR3\
         + sL*(21 - 21.2*sOL1*sOL2*sOR1*sOR2 - 3*sOL3*sOR3)\
         + sOL1*(-13.8 - RT*log(n)) + sOR1*(-12.7 - RT*log(n))\
         + sOL3*(-12.4 - RT*log(n)) + sOL2*(-12.1 - RT*log(n))\
         + sOR2*(-10.7 - RT*log(n)) + sOR3*(-10.2 - RT*log(n))'
```

Transcription at the PRM promoter is given by F_rm, which leads to the average AF_rm. There is also another promoter, PR, with transcription given by F_r, which leads to the average AF_r.

```
In [9]: F_rm='(45 + (415*(1 - sL) + 195*sL)*sOR2)*(1 - sOR3)'
        AF_rm=AveConf(F_rm,DG,S)
        AF_rm_n=simplify(subs(AF_rm,'RT',0.6))

        print AF_rm_n

(3.47792440482179e+51*n**5 + 3.97398116562347e+42*n**4 + 1.20593783102099e+32*n**3 +
2.51870689446793e+22*n**2 + 602508537514.617*n + 45.0)/(2.19946527169064e+58*n**6 +
1.08431513563455e+49*n**5 + 1.23153434343879e+40*n**4 + 1.15044168382098e+30*n**3 +
4.35928219885975e+20*n**2 + 12900662785.7089*n + 1.0)
```

```
In [10]: F_r='1200*(1 - sOR1)'
         AF_r=AveConf(F_r,DG,S)
         AF_r_n=simplify(subs(AF_r,'RT',0.6))

         print AF_r_n

(8.1395965296285e+49*n**5 + 1.29263057525327e+41*n**4 + 4.49899277535116e+32*n**3 +
4.86589652202721e+23*n**2 + 13611199302151.3*n + 1200.0)/(2.19946527169064e+58*n**6
+ 1.08431513563455e+49*n**5 + 1.23153434343879e+40*n**4 + 1.15044168382098e+30*n**3
+ 4.35928219885975e+20*n**2 + 12900662785.7089*n + 1.0)
```

Note that in both cases it is possible to obtain an exact analytical expression for the effective transcription rates, AF_rm and AF_r, as a function of the CI dimer concentration. The expression NF below gives the CI dimer concentration as function of the normalized CI monomer concentration.

```
In [11]: def NF(nn):
```

```
       return 9.38419e-14 + 7.0922e-10*nn - 1.15373e-11*sqrt(0.0000661586 + 1.*nn)
```

In [12]:
```
nnvals=arange(0, 2.5, 0.1)
nvals=array([NF(nn) for nn in nnvals])
```

Plotted below is AF_rm (solid red line) as a function of the normalized CI monomer concentration. The filled circles correspond the experimentally measured activity of the PRM promoter (Dodd, I.B., Shearwin, K.E., Perkins, A.J., Burr, T., Hochschild, A.and Egan, J.B.(2004) Cooperativity in longrange gene regulation by the lambda CI repressor, Genes Dev, 18, 344-354).

In [13]:
```
PRMdata = array([[0.058, 66.367], [0.373, 189.358], [0.658, 214.233],
                 [0.922, 188.897], [1.168, 173.696], [1.385, 155.271],
                 [1.875, 133.160], [2.281, 136.384]])

AF_rm_f=pythonFunction(AF_rm_n,'n')
plot(nvals, AF_rm_f(nvals), 'r--', PRMdata[:,0], PRMdata[:,1],'ob');
xlabel('$[CI]$',fontsize=14); ylabel('$AF_{rm}$',fontsize=14); show()
```

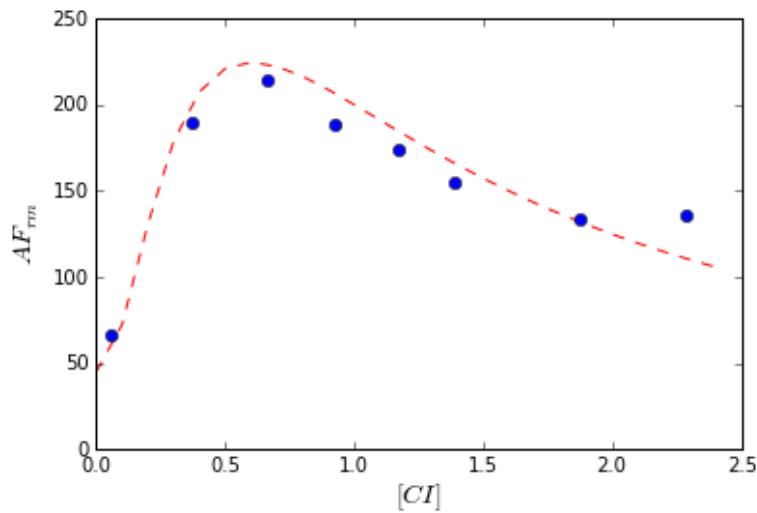

Plotted below is AF_r (solid red line) as a function of the normalized CI monomer concentration. The filled circles correspond the experimentally measured activity of the PR promoter ( Dodd, I.B., Shearwin, K.E., Perkins, A.J., Burr, T., Hochschild, A.and Egan, J.B.(2004) Cooperativity in longrange gene regulation by the lambda CI repressor, Genes Dev, 18, 344 - 354).

In [14]:
```
PRdata = array([[0.063, 1057.627], [0.373, 585.763], [0.658, 204.746],
                [0.921, 70.508], [1.171, 27.119], [1.384, 18.983],
                [1.875, 12.203], [2.278, 9.492]])

AF_r_f=pythonFunction(AF_r_n,'n')
plot(nvals,AF_r_f(nvals), 'r--', PRdata[:,0], PRdata[:,1],'ob');
xlabel('$[CI]$',fontsize=14); ylabel('$AF_{r}$',fontsize=14); show()
```

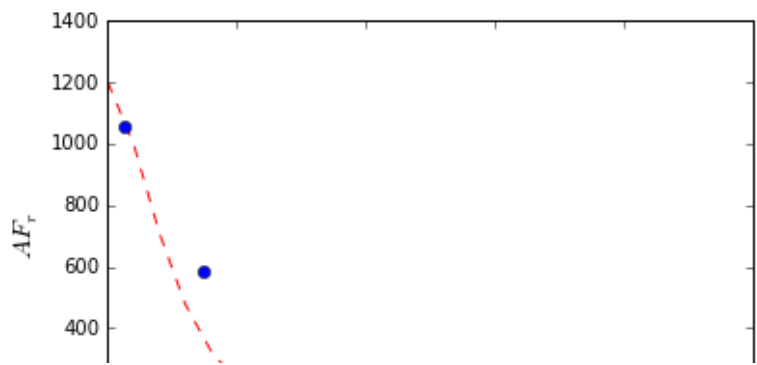

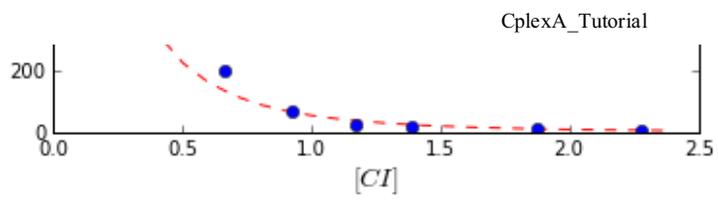